\documentclass[12pt,preprint]{aastex}

\newcommand{\PG} {PG 1115$+$166}
\newcommand{\logg} {\log g}
\newcommand{\Te} {T_{\rm eff}}

\newcommand{\msun} {$M_\odot$}
\newcommand\gta{\lower 0.5ex\hbox{$\buildrel > \over \sim\ $}} 
\newcommand\lta{\lower 0.5ex\hbox{$\buildrel < \over \sim\ $}} 
\newcommand{\nhe} {N({\rm He})/N({\rm H})}
\shortauthors{Bergeron et al.}
\shorttitle{Analysis of the DAB White Dwarf \PG}
\begin{document}


\title{Spectroscopic Analysis of the DAB White Dwarf \PG:\\
An Unresolved DA + DB Degenerate Binary}

\author{P. Bergeron}
\affil{D\'epartement de Physique, Universit\'e de Montr\'eal, C.P.~6128, 
Succ.~Centre-Ville, 
Montr\'eal, Qu\'ebec, Canada, H3C 3J7.}
\email{bergeron@astro.umontreal.ca}
\and
\author{James Liebert}
\affil{Steward Observatory}  
\affil{University of Arizona, Tucson AZ 85721}
\email{liebert@as.arizona.edu}

\begin{abstract}

A spectroscopic analysis of the DAB white dwarf \PG\ is presented. The
observed hydrogen and helium line profiles are shown to be
incompatible with model spectra calculated under the assumption of
homogeneous or stratified chemical compositions. In contrast, an
excellent fit to the optical spectrum of \PG\ can be achieved if the
object is interpreted as an unresolved double degenerate composed of a
hydrogen-line DA star and a helium-line DB star. The atmospheric
parameters obtained from the best fit are $\Te=22,090$~K and
$\logg=8.12$ for the DA star, $\Te=16,210$~K and $\logg=8.19$ for the
DB star. This binary interpretation is consistent with the results
recently reported by Burleigh et al. that \PG\ also exhibits radial
velocity variations. The implications of this discovery with respect
to the DAB spectral class are discussed.

\end{abstract}


\section{Introduction}

DAB stars are white dwarfs whose optical spectra are characterized by
the simultaneous presence of strong hydrogen Balmer lines together
with weaker neutral helium lines. To date, only a handful of these
objects are known \citep[see][and references therein]{burleigh}. DAB
stars are particularly important objects as they are likely to provide
the key to understanding the nature of the so-called ``DB gap'', a
range in effective temperature between $\Te\sim30,000$ and 45,000~K
where no helium atmosphere white dwarf (DO or DB stars) has ever
been identified \citep{liebert86}; the coolest DO star PG 1133$+$489 has an estimated
temperature of $\Te=46,000$~K
\citep{DW97}, while the hottest DB star PG 0112$+$104 is below
$\Te=31,500$~K \citep{beauchamp99}. The most widely accepted
explanation for the presence of this gap is the existence of competing
mechanisms that are affecting the atmospheric composition of white
dwarfs as they evolve along the cooling sequence \citep[see,
e.g.,][]{FW97}. Hence it is believed that residual hydrogen present in
the envelope of hot DO white dwarfs slowly diffuses towards the
surface, building an atmosphere that is gradually enriched with
hydrogen. By the time cooling DO stars have reached $\Te\sim45,000$~K,
they would all bear the signature of a hydrogen-rich atmosphere, a DA
star. Such DA stars should have a thin hydrogen atmosphere in
diffusive equilibrium on top of the helium envelope. At the red edge
of the DB gap, convective dilution of this thin hydrogen atmosphere
with the underlying convective helium envelope is then believed to
turn $\sim20$\% of all DA stars near $\Te\sim30,000$~K into DB white
dwarfs. The identification of white dwarfs with hybrid spectra in
these particular ranges of effective temperature has always been of
significant interest since they may represent objects transitioning
from one spectral type to another.

Recently, \citet{burleigh} have reported the discovery of a new DAB
white dwarf, \PG, as part of a survey aimed at identifying helium-rich
objects crossing the DB gap. The authors have examined hot white
dwarfs that went undetected by the ROSAT and EUVE surveys of the
extreme ultraviolet sky. Effective temperatures and surface gravities
for the objects in their sample had been supplied by us as part of
another project aimed at redetermining the luminosity function of DA
stars using $\Te$ and $\logg$ values obtained from detailed fits to
the hydrogen lines \citep[see, e.g.,][]{BSL}. As such, the DAB nature
of \PG\ was already known to us back in 1996 from our own
spectroscopic observations of the PG sample. The optical spectrum of
\PG\ had also been thoroughly analyzed using model atmospheres with
homogeneous and stratified helium/hydrogen compositions. In this
paper, we report the results of our findings. Interestingly enough, we
show that the conclusions of \citet{burleigh} that \PG\ is probably a
double degenerate binary composed a DA and a DB star is confirmed by
our detailed analysis.

We first present our spectroscopic observations and model atmosphere
calculations in \S\S\ 1 and 2, respectively. The optical spectrum is
then analyzed in detail in \S\ 3 using using homogeneous,
stratified, and composite model atmospheres. Our discussion of the
results follows in \S\ 4.

\section{Spectroscopic Observations}

Optical spectroscopy for \PG\ has been obtained on 1996 May 20 using
the Steward Observatory 2.3-m reflector telescope equipped with the
Boller \& Chivens spectrograph and a Loral CCD detector. The 4.5
arcsec slit together with the 600 l/mm grating in first order provided
a spectral coverage of $\lambda\lambda$3200--5300 at an intermediate
resolution of $\sim 6$~\AA\ FWHM.

Our optical spectrum for \PG\ is compared in Figure
\ref{fg:f1} with those of MCT 0453$-$2933 and MCT 0128$-$3846 
taken from \citet{wes94}. There is an obvious similarity between all
three objects. The HeI $\lambda\lambda$4026 and 4471 absorption lines
can be seen in all three spectra with about equal shape and
strength. Weaker HeI features at $\lambda\lambda$4713, 4921,
and 5015 are also observed in \PG\ and MCT 0453$-$2933; these lines
are much weaker but nevertheless present in the spectrum of MCT
0128$-$3846. The two MCT objects have been analyzed by \citet{wes94}
and shown to be unresolved, composite systems consisting of a DA white
dwarf together with a DB or DBA star.

\section{Model Atmospheres and Synthetic Spectra}

The mixed hydrogen and helium model atmospheres and synthetic spectra
used in this analysis are described in \citet{dao94}. These are
calculated assuming either a homogeneous or a stratified chemical H/He
composition. The homogeneous model atmospheres are similar to those
described in \citet{wes80}, while the stratified configurations have
been calculated using the formalism discussed at length in
\citet{jordan86} and \citet{vennes92}. The 
homogeneous model grid covers a range of
$\Te=20,000~(5000)~100,000$~K, $\logg=6.5~(0.5)~8.5$, and
$\log\nhe=-5.0~(1.0)~0.0$ (where the numbers in parentheses indicate
the step value), as well as pure hydrogen models. Stratified models have
been calculated for the same range of effective temperature and
surface gravity, with $\log q_{\rm H}\equiv \log M_{\rm
H}/M_{\star}=-17.0~(0.5)-15.0$ at $\logg=8.0$.  As discussed in
\citet{dao94}, the particular range of $\log q_{\rm H}$ used varies with
$\logg$ as to yield synthetic spectra that look relatively similar
with respect to the strength of the helium lines in the $\logg$ --
$\log q_{\rm H}$ plane.

The synthetic spectra for the mixed H/He models have been calculated
following \citet[][see also \citealt{BSL,bergeron93}]{bergeron91}, where
the treatment of the hydrogen line profiles is described at length.
The HeI lines have been treated as Voigt profiles with the Stark
broadening parameters listed in \citet{dao94}.

Our grid of model atmospheres and synthetic spectra for DB stars is
described in \citet{beauchamp96}. These include the improved Stark
profiles of neutral helium of \citet{beauchamp97}. The model
atmospheres used here assume a pure helium composition and cover a
range of $\Te=10,000~(1000)~16,000~(2000)~30,000$~K and
$\logg=7.0~(0.5)~9.0$.

\section{Model Atmosphere Analysis}

Our fitting technique relies on the nonlinear least-squares method of
Levenberg-Marquardt \citep{press86}, which is based on a steepest
descent method. The model spectra (convolved with a Gaussian
instrumental profile) and the optical spectrum of \PG\ are first
normalized to a continuum set to unity. The calculation of $\chi ^2$
is then carried out in terms of these normalized line profiles
only. All atmospheric parameters -- $\Te$, $\logg$, $\nhe$ or $\log
q_{\rm H}$ -- are considered free parameters. When fitting DA + DB
model spectra, the total flux of the system is obtained from the sum
of the monochromatic Eddington fluxes of the individual components,
weighted by their respective radius. Those radii are obtained from
the thin and thick hydrogen envelope evolutionary models with
carbon/oxygen cores described in \citet{BLR}.

Our best fits to the optical spectrum of \PG\ using homogeneous,
stratified, and composite DA+DB models are shown in Figure
\ref{fg:f2}. The solution using homogeneous models, $\Te=32,000$~K, 
$\logg=7.95$, and $\log \nhe=-1.44$, is in excellent agreement with
that obtained by \citet{burleigh}, $\Te=33,000$~K and $\log \nhe=-1.5$
(their value of $\logg$ is not given). The problem mentioned by these
authors with the HeII $\lambda$4686 predicted by their models is not
observed here, however, even though this particular line is included
in our model calculations. The fits to the hydrogen Balmer lines are
qualitatively similar between the homogeneous and stratified models:
the low Balmer lines (H$\beta$ to H$\delta$) are predicted too strong
while the higher members are too weak.  The homogeneous model yields a
satisfactory fit for the HeI lines $\lambda\lambda$4026, 4713, and
5015, but a poor fit to the remaining helium lines. The fit with the
stratified model is considerably worse.

On the other hand, the DA+DB solution provides an excellent fit to the
Balmer and neutral helium lines simultaneously. All observed features
are reproduced in detail and this is clearly the only viable solution
for \PG. The effective temperatures determined for both the DA and the
DB components are significantly lower than the values achieved under
the assumption of a single object with a homogeneous or stratified
composition. This is a direct consequence of the hydrogen lines being
diluted by the continuum flux of the DB star (see below); the Balmer lines
are weakened, and the effective temperature of the model needs to be
increased to match the observed line profiles.

The surface gravities of both components of the system are almost
identical, $\logg\sim 8.15$, which corresponds to a mass of $\sim0.7$
\msun\ using the evolutionary models described above. The white dwarf
cooling ages inferred for the DA and DB stars are respectively
$6.0\times10^7$ and $2.2\times10^8$ years. Since both stars have
roughly the same radius, the contribution of each component to the
total luminosity is only a function of the effective
temperature. Since the DB star is significantly cooler than the DA
component, the former will contribute less to the combined luminosity
of the system. This is illustrated more quantitatively in Figure
\ref{fg:f3} where the contribution of each component to the total
flux is depicted. Also shown is the multichannel $ugvr$ photometry
taken from \citet[][$v=15.14$, $u-v=0.15$, $g-v=-0.14$,
$g-r=-0.51$]{green86}, which is in good agreement with our adopted
solution, with the exception of the $u$ magnitude; multichannel $u$
magnitudes were sometimes compromised by the atmospheric dispersion,
the finite size of the entrance aperture, and the drift of the target
away from the aperture center during the exposure. While the flux in
the ultraviolet is completely dominated by the DA component, there is
a more significant contribution of the DB star in the optical regions
of the energy distribution. In particular, the cores of the lower
Balmer lines are filled in by the continuum flux of the DB star,
resulting in the poor fits of the homogeneous and stratified solutions
displayed in Figure \ref{fg:f2} and discussed above.

Finally, \citet{burleigh} have reported radial velocity variations in
\PG\ from H$\alpha$ measurements (see their Fig.~2), with possible HeI
$\lambda$6678 moving in anti-phase with the hydrogen line, in
agreement with a DA+DB interpretation. No orbital period has yet been
reported.

\section{Discussion}

Our analysis has shown conclusively that the simultaneous presence of
hydrogen and helium in the spectrum of \PG\ is a simple result of an
unresolved binary system composed of a DA white dwarf and a DB
star. Both components of the \PG\ system lie well below the cool edge
of the DB gap, and as such, this DAB ``system'' cannot contribute to
our understanding of the nature of the DB gap. Two other DAB stars,
MCT 0128$-$3846 and MCT 0453$-$2933 shown in Figure \ref{fg:f1}, are
similar DA + DB (or DBA) unresolved composite systems, rather than
single objects with mixed hydrogen and helium compositions.

Additional DAB stars include HS 0209$+$0832, a $\Te\sim35,000$~K white
dwarf that has been recently observed with HST and analyzed by
\citet{wolff00}. The authors conclude that the hydrogen and helium lines,
as well as numerous metallic features, are all compatible with a
homogeneously mixed composition, and that the object is probably in the
process of accreting matter from the interstellar medium. If this is
the case, HS 0209$+$0832 is a transient phenomenon, and is not
connected to the DB gap either. Its temperature within the DB gap is
thus purely coincidental.

The case of G104$-$27 reported by \citet{holberg90}, with a
temperature of $\Te=26,000$~K, is also peculiar. Their spectrum shows
a weak HeI $\lambda$4471 feature, and possible
$\lambda$5015. However, subsequent observations of this object,
including 7 independent spectra collected over the years by one of us
(PB), reveal a featureless spectrum near the $\lambda$4471
region. This particular spectroscopic feature may of course be
variable, as is the case for HS 0209$+$0832 \citep{edelmann}. Such
variations could be expected if G104$-$27 and HS 0209$-$0832 accrete
from an inhomogeneous interstellar medium, or if the distribution of
helium on the stellar surface is inhomogeneous, as discussed by
\citet{edelmann}.

Finally, there is GD 323, the prototype of DAB
stars. \citet{liebert84} showed that the combined ultraviolet and
optical spectrophotometry of this star could not be reproduced with a
homogeneously mixed hydrogen and helium composition, and suggested
that a fit with a stratified composition be attempted.  This idea was
explored more quantitatively by \citet{koester94} who confirmed that
the most promising explanation to account for the observed line
profiles was a stratified atmosphere. Their estimated temperature of
$\Te=28,750$~K places GD 323 just below the cool end of the empirical
DB gap. However, even their stratified solution does not reproduce the
optical spectrum very well, at least not at the level of our best fit
for \PG\ shown in Figure
\ref{fg:f1}, and inconsistent values of the hydrogen layer mass are
derived from the optical and ultraviolet spectra. 

All in all, there seems to be no such objects as ``DAB stars
transiting the DB gap'', as sought by \citet{burleigh}; we mention
also that no additional DAB spectrum has been identified in our PG
luminosity function sample. This is perhaps not surprising,
however. Indeed, if the scenario proposed to explain the DB gap is
correct, by definition we should not observe any helium-rich objects
crossing the DB gap. Stars with mixed hydrogen and helium compositions
must be searched either slightly above the hot end of the DB gap
($\Te\
\gta45,000$~K) or slightly below the cool end ($\Te\
\lta30,000$~K). Interestingly enough, such stars have probably already
been identified. PG 1305$-$017 analyzed by
\citet{dao94}, with an estimated temperature of $\Te=44,400$~K, is the perfect 
example of a hybrid white dwarf with strong hydrogen and helium lines
-- a DAO star in this case -- that can be successfully explained in
terms of a stratified chemical composition (see Fig.~8 of Bergeron et
al.). This object corresponds precisely to our expectation of a DO
star turning into a DA star near the hot end of the DB gap, i.e. a
thin hydrogen atmosphere in diffusive equilibrium on top of a helium
envelope. Similarly, at the cool end of the DB gap, the reappearance
of helium-rich atmosphere white dwarfs near 30,000~K is explained in
terms of the dilution of this thin hydrogen atmosphere with the more
massive underlying helium convection zone. It is clear that DA white
dwarfs caught in the process of being convectively diluted will have a
complex chemical stratification that will not be homogeneous nor
stratified. GD 323 with a temperature slightly below the cool edge of
the DB gap might be such an exotic object. Hence, objects that could
support the DO to DA and DA to DB transition scenarios do exist, but
they should not be looked for within the DB gap itself. This
interpretation is at least currently supported by observations.

One possible scenario for the origin of Type Ia Supernovae is
the double degenerate model in which two white dwarfs with a total
mass exceeding the Chandrasekhar limit are assumed to coalesce
\citep{webbink84,iben84}. In order for the system to merge within a
Hubble time, however, the orbital period must be less than $\sim 10$
hours. Since the total mass of the \PG\ system ($M\sim1.4$ \msun) is
close to the Chandrasekhar limit, a precise measurement of its orbital
period would establish whether \PG\ represents a likely supernova
candidate.

\acknowledgements This work was supported in part by the NSERC
Canada, by the Fund FCAR (Qu\'ebec), and by the NSF grant AST
92-27961.

\clearpage
\figcaption[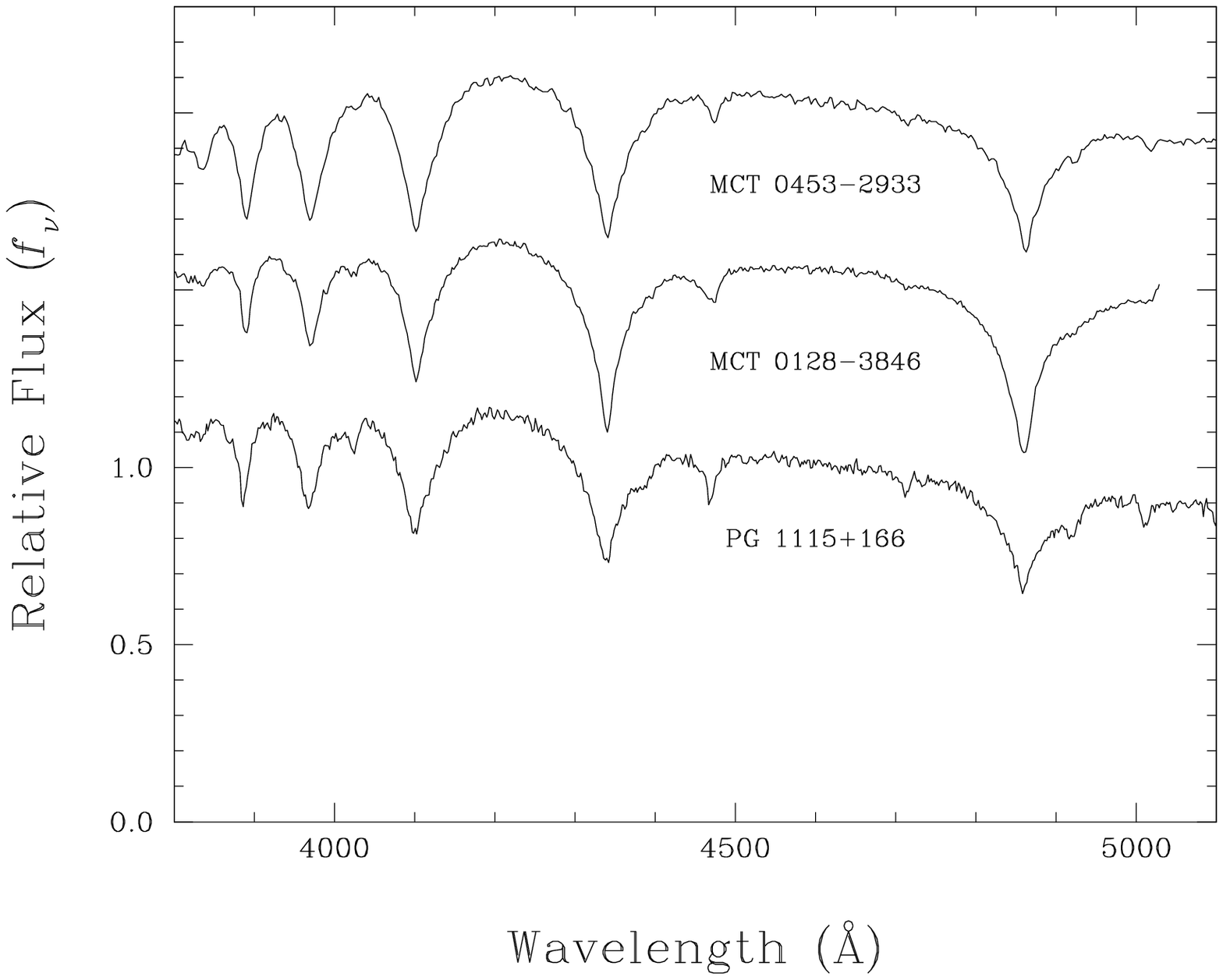] {Comparison of our optical spectrum of the DAB 
star \PG\ with those of MCT 0453$-$2933 and MCT 0128$-$3846,
two additional DAB stars that have been interpreted by \citet{wes94}
as unresolved composite systems consisting of a DA white dwarf and a
DB or DBA companion. The spectra are normalized at 4400 \AA\ and are
shifted vertically by 0.5 for clarity.\label{fg:f1}}

\figcaption[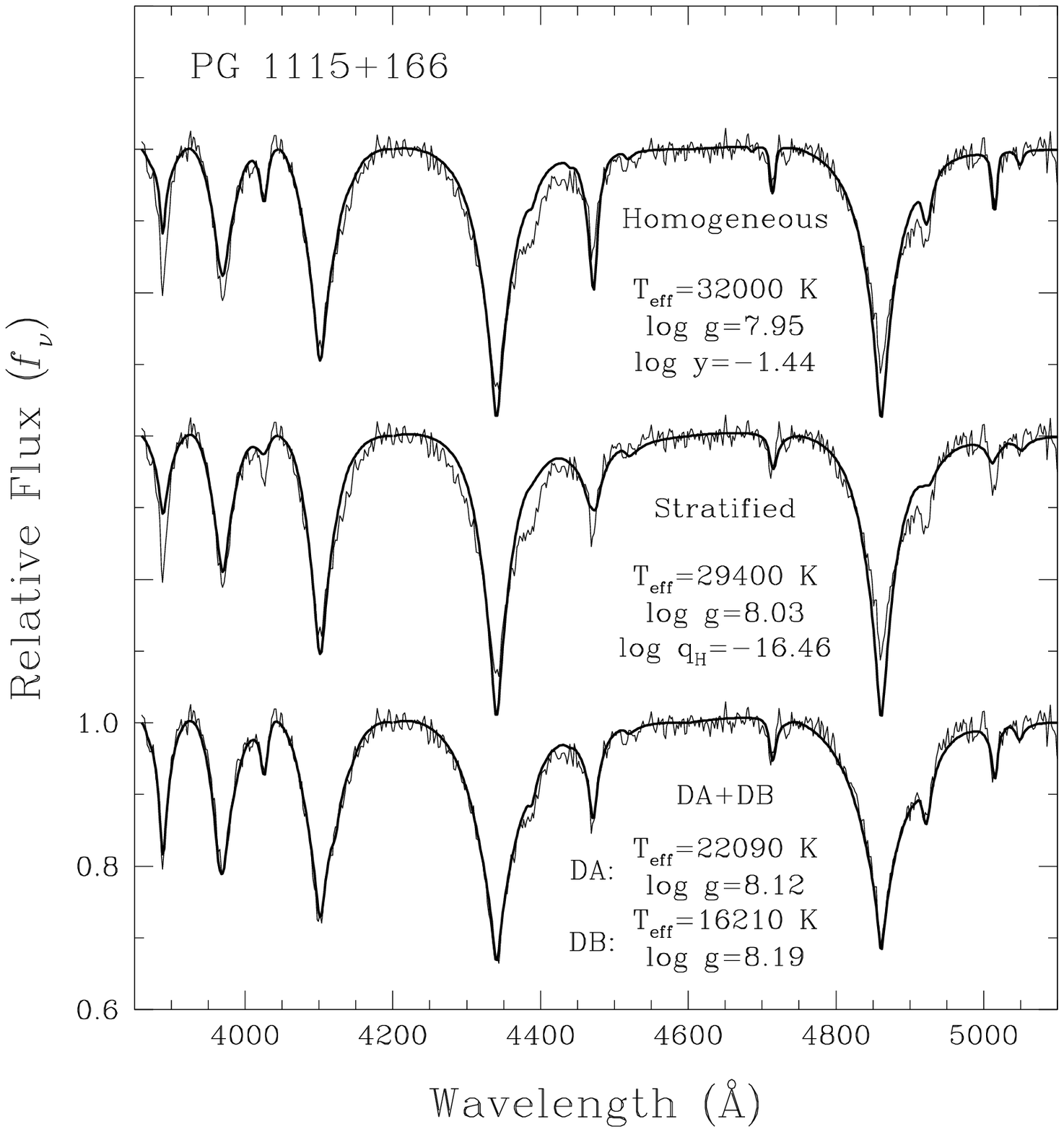] {Our best fits to the optical spectrum of 
\PG\ using homogeneous, stratified, and composite DA+DB
models. The atmospheric parameters for each solution are given in the
figure. Both the observed and theoretical spectra are normalized to a
continuum set to unity. The top spectra are shifted by a factor of 0.4
from each other for clarity. Clearly, the DA+DB solution provides the
best fit to the overall spectrum.\label{fg:f2}}

\figcaption[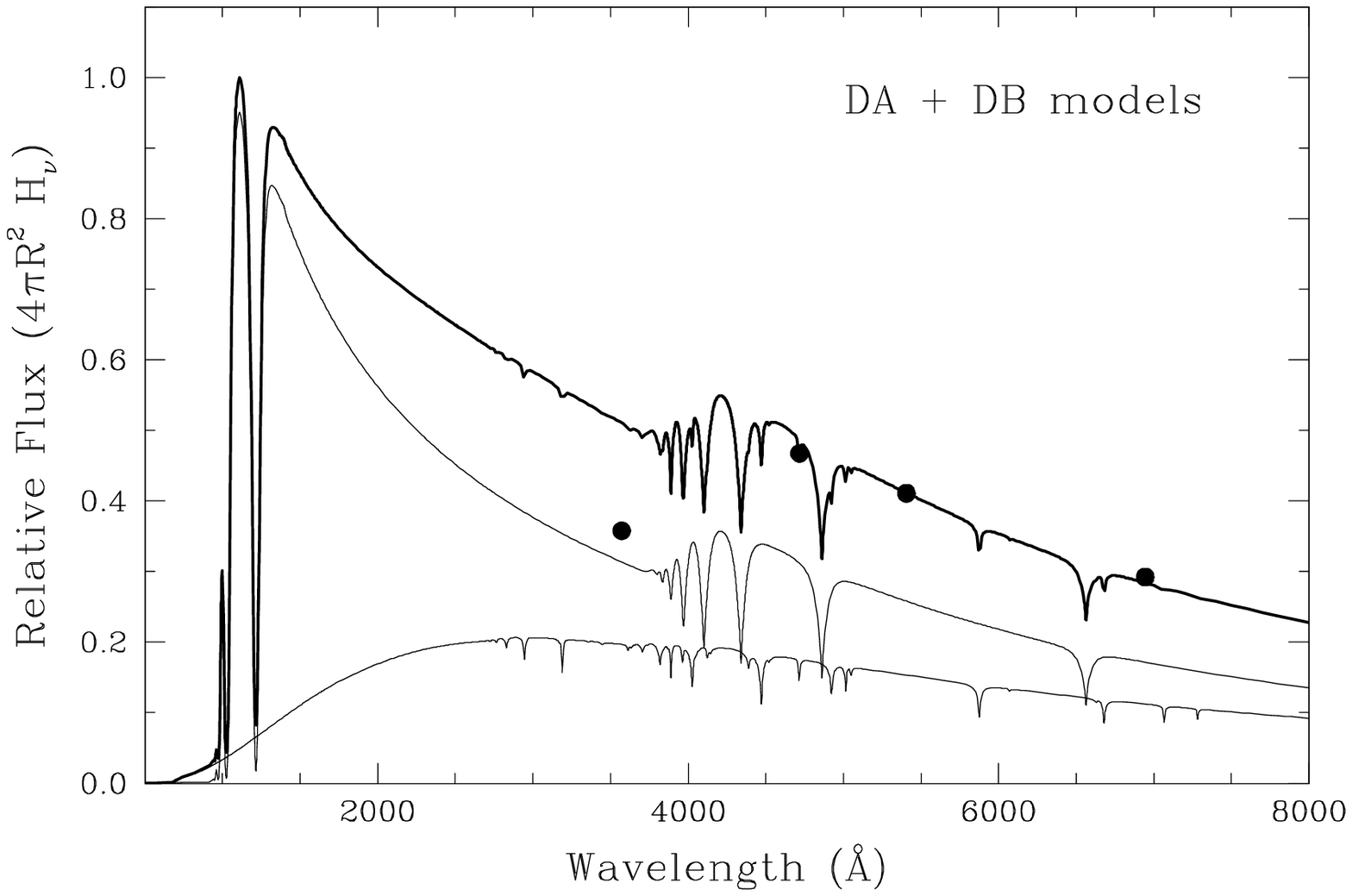] {Relative energy distributions for our best DA+DB fit
obtained in Figure \ref{fg:f2}. The thin lines show the individual
contributions of the DA and DB components, properly weighted by their
respective radius, while the thick line corresponds to the total
monochromatic flux of the composite system. The dots indicate the
multichannel $ugvr$ photometry normalized at the $v$
magnitude to match the predicted energy distribution.\label{fg:f3}}

\clearpage
\begin{figure}[p]
\plotone{f1.ps}
\begin{flushright}
Figure \ref{fg:f1}
\end{flushright}
\end{figure}

\begin{figure}[p]
\plotone{f2.ps}
\begin{flushright}
Figure \ref{fg:f2}
\end{flushright}
\end{figure}

\begin{figure}[p]
\plotone{f3.ps}
\begin{flushright}
Figure \ref{fg:f3}
\end{flushright}
\end{figure}

\end{document}